\documentclass[conference]{IEEEtran}
\IEEEoverridecommandlockouts
\usepackage{flushend}
\usepackage{cite}
\usepackage{amsmath,amssymb,amsfonts}
\usepackage{algorithmic}
\usepackage{graphicx}
\usepackage{textcomp}
\usepackage{xcolor}
\usepackage{amsthm}
\def\BibTeX{{\rm B\kern-.05em{\sc i\kern-.025em b}\kern-.08em
    T\kern-.1667em\lower.7ex\hbox{E}\kern-.125emX}}

\usepackage{hyperref}
\usepackage{balance}
\usepackage{xcolor}
\usepackage{soul}

\usepackage{diagbox}
\usepackage{graphicx}
\usepackage{multirow}
\usepackage[utf8]{inputenc}
\usepackage{subcaption}
\usepackage{enumitem}

\usepackage{amsmath}%
\usepackage{wasysym}%

\usepackage{tikz}
\usepackage{xargs}                      % Use more than one optional parameter in a new commands
\newlength{\leftLength}
\usepackage{colortbl}

\usepackage{color}
\usepackage{soul}

\definecolor{lightgrey}{rgb}{0.925, 0.925, 0.925}
\sethlcolor{lightgrey}

\makeatletter
\def\SOUL@hlpreamble{%
    \setul{}{3.5ex}% increased by 1ex
    \let\SOUL@stcolor\SOUL@hlcolor
    \dimen@\SOUL@ulthickness
    \dimen@i=-.75ex % increased by -0.25ex
    \advance\dimen@i-.5\dimen@
    \edef\SOUL@uldepth{\the\dimen@i}%
    \let\SOUL@ulcolor\SOUL@stcolor
    \SOUL@ulpreamble
}
\makeatother

\newcommand{\summary}[2]{
\setlength{\leftLength}{0.5cm + (\widthof{{\small \textbf{#1}}}/2)}
\vspace{2ex}\noindent\begin{tikzpicture}
\node[align=center,draw,thin,minimum width=\columnwidth,inner sep=2.2mm] (titlebox)%
{\parbox{0.95\columnwidth}{\vspace*{0.25ex}\noindent\emph{#2}}};\node[fill=white] (W) at ([xshift=\the\leftLength] titlebox.north west) {{\small \textbf{#1}}};%
\end{tikzpicture}\vspace{2ex}}

\newcommandx{\email}[1]{{\href{mailto:#1}{#1}}}

\begin{document}

\title{The complexity paradox:  An analysis of modeling education through the lens of complexity science}

\author{
    \IEEEauthorblockN{Daniel Strüber}
    \IEEEauthorblockA{\textit{Chalmers $\mid$ University of Gothenburg (SE)}, \textit{Radboud University Nijmegen (NL)}}
    %\\\email{danstru@chalmers.se}\vspace{-15pt}}
}

\maketitle

\begin{abstract}
Modeling seeks to tame complexity during software development, by supporting design, analysis, and stakeholder communication. Paradoxically, experiences made by educators indicate that students often perceive modeling as adding complexity, instead of reducing it. In this position paper, I analyse modeling education from the lens of \textit{complexity science, }a theoretical framework for the study of complex systems.
I revisit pedagogical literature where complexity science has been used as a framework for general education and subject-specific education in disciplines such as medicine, project management, and sustainability. I revisit complexity-related challenges from modeling education literature, discuss them in the light of complexity and present recommendations for taming complexity when teaching modeling.
\end{abstract}

% \begin{IEEEkeywords}
% software product lines, modeling
% \end{IEEEkeywords}

%\section{Introduction} \input{intro}~\newpage~\newpage~\newpage

\section{Introduction}
\label{sec:intro}
\label{sec:patternsbyexample}

% Software systems are among the most intricate artifacts created by humans. A single software system can consist of several million lines of code, written by thousands of developers in a large-scale organization or open-source software project. The enormous magnitude of software systems arises because m
Modern software is inherently complex: it typically needs to address a huge variety of interactions with users, hardware devices, physical application contexts, and other software systems, typically in highly dynamic and continuously changing real-world environments.
As a tool to deal with complexity during software development, companies rely on \textit{modeling}.
Brook’s famous essay on essential and accidental complexity  \cite{brooks1987no} 
offers a framework for motivating the use of models, as a tool that abstracts from the accidental
complexity involved in low-level programming, and allows developers to focus on the unavoidable,
essential complexity of the problem at hand.

Paradoxically, experiences in software engineering education indicate that students often perceive modeling as adding, instead of removing, complexity during software development, and struggle with the associated challenges \cite{brosch2009teaching}.
Modeling is usually taught in undergraduate courses, either in dedicated courses on software design, or as a section in broader introductory courses on software engineering.
In these courses, students typically engage with small modeling tasks.
In such contexts, explicitly designing software systems in a new language presents a radical departure from the programming activities that students are familiar with, and students often struggle with using dedicated tools that they perceive as complex and not particularly user-friendly.

In this position paper, I analyse modeling education from the lens of the theoretical framework of complexity science \cite{davis2005challenging,davis2006complexity}. Complexity science is involved with the study of systems that are highly dynamic, uncertain and non-linear in nature.
It has previously been used as a framework to problematize and suggest improvements to educational practice, both in general \cite{davis2005challenging,davis2006complexity} and subject-specific education \cite{gormley2016learning,thomas2008preparing,weber2021teaching}.
As such, complexity science provides a vocabulary that allows exploring connections between challenges in modeling education and the previous pedagogical literature, and drawing inspiration from other disciplines' ways of dealing with complexity.
I provide an overview of background in this direction in Sect.~\ref{sec:background}.

%between complexity-related issues in modeling education and previous pedagogical literature, where complexity science has been used as a framework for 

To analyse complexity-related challenges in modeling education, I apply a lens of interpretation based on the distinction of two mindsets---an \textit{analytical} one and a \textit{complexivist} one--, lifted from educational literature. I motivate and explain this approach in Sect.~\ref{sec:framework}.
Through this lens of interpretation, in Sect.~\ref{sec:challenges}, I revisit recurring challenges reported in modeling education literature and discuss them in the light of complexity science, with a focus on misalignments between analytical-mindset teaching practices and the complex situations that students will encounter in practice, which require a complexitivist mindset.
 To address the identified challenges, in Sect.~\ref{sec:recommendations}, I present recommendations for addressing complexity.

%As the distinguishing feature of this analysis, complexity science provides a vocabulary that allows exploring connections to pedagogical literature from other fields, and draw inspiration from their ways of dealing with complexity in education.

\section{Background}
\label{sec:background}

I now revisit complexity science and its use as a framework for discussing education. In particular,  I describe how complexity science  has previously been employed as a framework for problematizing aspects of higher education.

\smallskip
\noindent{}\textbf{Complexity science}  \cite{zimmerman1998complexity,manson2001simplifying,phelan2001complexity} deals with systems that are dynamic, uncertain and non-linear, the so-called \textit{complex systems}. %Complexity science  has seen broad application among  hard and soft sciences. 
%Since complexity science is defined by its object of study, rather than a particular method, approach, or world-view, there is no consensus on the nature of what complexity theory is---e.g. a theory, a discipline, or a paradigm \cite{davis2006complexity}.
Complex systems have several qualities that make them particularly complicated. They are usually made up by independent actors who show some sense of \textit{self-organization}. The behavior of the constituent actors leads to phenomena of \textit{emergence}: observations that cannot be explained only by the behavior of individual constituent elements. The actors in such systems are defined by their relationships, where \textit{short-range relationships} are vital for determining the system's overall behavior. 
The structure of a complex system continuously evolves, which allows it learn over time, especially to respond to \textit{pertubations} (e.g., disruptions, stress).
%Complex systems are usually far from a stable equilibrium state, since an equilibrium would lead to idleness and, therefore, the death of the system 
%Complex systems are \textit{organizationally closed }in the sense that they endure over time, while they are \textit{open to their surroundings }and may integrate and disintegrate actors over time  \cite{davis2005challenging,davis2006complexity}.

%Complex systems are distinguished from less complex systems, including \textit{simple systems} with basic cause-effect relationships that can be described by linear, analytical methods (e.g., Newton-style mechanics), and \textit{complicated systems} involving a large number of interconnected parts governed by statistical laws, which yet do not show emergence (e.g., astronomical phenomena).

%Complexity science  has begun to influence higher education literature in the last 20 years, both in general and field-specific education literature. Concrete contributions have been made to position complexity science as a theoretical framework for explaining education, and in practical applications of such a framework in specific disciplines. I now discuss work in both directions.

\smallskip
\noindent{}\textbf{Complexity as a framework for explaining education. }Davis and Sumara have presented a positioning of complexity science as a theory for education \cite{davis2005challenging,davis2006complexity}. Their positioning challenges traditional views on education that emphasize dichotomies, e.g., viewing knowledge and knowers as entities that exist independently from each other, which they contrast with the notion of \textit{simultaneity}, referring to phenomena that exist or operate at the same time.
In this context, they emphasize the view of the classroom as a complex \textit{learning system}  which brings forward knowledge as an emergent phenomenon, arising from teachers and learners in their unique context, including personal, societal, cultural, and ecological levels.
%A crucial notion  is that of \textit{level-jumping}--``\textit{that is, [to] simultaneously
%examine [a] phenomenon in its own right (for its particular coherence
%and its specific rules of behavior) and pay attention to the conditions of its
%emergence (e.g., the agents that come together, the contexts of their co-activity,
%etc.)}'' %\cite{davis2006complexity}.
%In this view, effective teachers are able to level-jump, as they are aware of the interconnected levels of the learning situation at hand, and can design learning activities in a way that is conducive to learning as a form of emergence.

%As an important distinction, which is fundamental for our analysis, they mindsets: 

%\smallskip
%\noindent{}\textbf{Analytical and complexivist mindsets.}
From their theoretical framework, Davis and Sumara derive consequences for teaching methods.
A key distinction is  between two mindsets, the \textit{analytical} and the \textit{complexivist} one \cite{davis2010if}.
The analytical mindset, which is advocated by traditional teaching methods,  is focused on linear, analytical knowledge,  teaching students skills that are suited for systems of limited complexity.
A key tenet of the analytical mindset is \textit{reductionism}, which seeks to remove complications from the analysis of situations to reduce them to the essential elements.
%Yet, in a world in which knowledge itself is getting increasingly uncertain and contestable, 
%the analytical mindset learning activities are deemed no longer sufficient in this view. 
%In a complexivist mindset, this is not suitable to 
A criticism is that this mindset might not be adequate to prepare students to real-world situations, in which they are faced with the complications inherent to complex systems.
To foster the capacity for dealing with complexity, they propose to engage with learning activities that are conducive to a mindset that embraces complexity, a so-called complexivist mindset.
%With the elements of complex systems in mind, such learning activities incorporate aspects that foster an attentiveness to context, connection, and contingency.
%Below I discuss selected examples for such learning activities.

\smallskip
\noindent{}\textbf{Using complexity science to guide the design of learning activities}. The abovementioned  educational framework has inspired contributions that apply complexity science to develop learning activities in specific disciplines. I now discuss selected examples from four disciplines.

In computing education, Fabricatore et al. \cite{fabricatore2014complexity} adopted a complexity-science-based  teaching strategy in a game design course. Towards the course goal of developing a game for a museum, the course included three project milestones, project workshops, lectures, and formative tasks. The course was designed in an adaptive way, by considering the emerging evidence as the course was given. Applied complexity principles included decentralised control and self-organisation within teams, frequent and unexpected perturbations (in the form of consultant feedback) triggering adaptive dynamics, selection of course contents based on external, contextual factors, and iterative, incremental teaching applying concepts at increasing levels of depth and complexity.

In medical education, educators need to prepare students for multi-faceted clinical situations with complicating factors such as time pressure, sensitive human aspects, and ethical dilemmas. Gormley and Fenwick present a \textit{ward-based simulation} exercise \cite{gormley2016learning}, in which students interact with actors during a realistic ward situation. The situation is made increasingly more complex, to include aspects of inappropriate patient and colleague behavior, and life-and-death-related decision making. Data collected during a run of the simulation allowed the uncovering of complexity-related patterns.
They derive a set of strategies for managing complexity, namely: taking time to size up the system; attuning to what emerges; reducing complexity; boundary practices; and working with uncertainty.

%Challenges regarding complexity were: being unprepared for 'diving in', being caught in an escalating system, being bounded by the present of a patient, being unable to assert boundaries of acceptable practices.

 Project management education has traditionally been informed by professional standards such as PMBOK and APM, describing formalized, context-independent processes \cite{thomas2008preparing}. To better prepare project managers for complex projects with sudden changes, disruptions and interpersonal challenges,  Thomas and Mengel propose a re-shift of project manager education towards the development of ``emotionally and spiritually intelligent'' project leaders, who comfortably navigate complex systems in the transition between a stable and a perturbed state \cite{thomas2008preparing}. They suggest project-based learning modes that take complexity science seriously, fostering an \textit{ability to adapt to change} and to \textit{develop new approaches on the fly}.

 In sustainability education, it has been long recognized that sustainability problems are wicked problems \cite{weber2021teaching} that cannot be addressed in terms of isolated factors, as they typically involve an interplay of complex societal and ecological circumstances. Complexity science offers mathematical methods for describing complex phenomena that can be used to support an explicit analysis of such factors in educational activities. Weber et al. \cite{weber2021teaching} report on a workshop in which participants from diverse backgrounds were prompted to prepare graph-based system representations of a selected sustainability problem, which were then discussed in-plenum. 
 Feedback
indicates that the activity furthered the participants' interdisciplinary thinking.

% In summary, complexity is a problem of interest across higher education.
% Previous literature has introduced  theoretical treatments and concrete examples for using complexity science in educational contexts.
% The examples guide my  own approach in this paper in a twofold way:
% first, as inspiration, showing that a shift to complexity-aware  activities can enable productive teaching and learning.
% Second, by contributing  teaching strategies that I will use to derive recommendations.

\section{Lens of interpretation}
\label{sec:tenets}
\label{sec:framework}

In this paper, I apply a complexity-science-based \textit{lens of interpretation} to discuss and inform the design of learning activities in modeling	education.
This lens comprises two basic tenets that build on the conception of analytical and complexitivist mindsets \cite{davis2010if}, as introduced earlier:
Methodologically, this approach is inspired by \textit{model papers} \cite{jaakkola2020designing}, a type of conceptual article that summarizes arguments about a focal construct as a set of logical statements, in this case, the tenets.
%Below, I first state the tenets, before explaining them further in the following text.

\smallskip
\newcommand{\framedtext}[1]{%
\par%
\noindent\fbox{%
    \parbox{\dimexpr\linewidth-2\fboxsep-5\fboxrule}{#1}%
}%
}

\framedtext{
\vspace{-2pt}
\begin{enumerate}
	\item 
Several of the major challenges in modeling education arise from a misalignment of mindsets:
At the time that students first engage with modeling, their previous education has guided them to develop an \textit{analytical mindset}. In contrast, problems of a type in which modeling becomes beneficial require the adoption of a \textit{complexivist mindset}.
\item Learning activities in modeling education should facilitate a shift between these mindsets. To that end, they should be explicitly designed with complexity in mind.
\end{enumerate} 
\vspace{-2pt}
}

\smallskip

Before taking a modeling course, students have typically participated in first-year programming courses, potentially in addition to high-school courses and hobby projects, where they dealt with problems that were linear in nature, facilitating the adoption of an \textit{analytical} mindset.
 Such problems have several key commonalities: First, they can be solved entirely by an reductionist approach, in which the overall problem is reduced to several smaller, sub-problems (the ``divide and conquer'' paradigm; see the standard examples in programming courses, such as \textit{searching} and \textit{sorting}). Second, these problems grow linearly in size, in terms of the number of entities in the system. For example, adding a filter in a ``pipes and filter'' architecture requires adding code at one specific location in the overall codebase, instead of modifications at many places. Third, problems are specified in a clearly delineated ``box'' (input and output specifications of an algorithm), abstracting away from real-world intricacies, and their inherent uncertainty. For such basic problems, students are well-prepared with the contents of first-year programming courses.

In contrast, the nature of complex real-world problems that software deals with, in which modeling unfolds its potential, require a complexivist mindset \cite{davis2010if} to cope with complexity.
Complexity arises because software systems are deployed to the real world, and as such, are prone to changes, 
require to make decisions in a certain context,
might be affected by uncertainty,
and involve people in various roles, who may or may not have shared understandings, perspectives, and interests.
Drawing a parallel to recent developments in other disciplines affected by such factors (discussed in Sect.~\ref{sec:background}), I make a case that modeling education could greatly benefit from a mindset shift:
To support the development of a mindset that prepares students for complexity, we need learning activities that explicitly take aspects of complex systems into account.

\section{Discussing  challenges in modeling education}
\label{sec:challenges}

I  now revisit recurring reported challenges retrieved from field-specific  literature
and discuss them from the chosen lens of interpretation.
The focus of this section is on a problem analysis, before describing solutions  in the next section.

\smallskip
\noindent{}\textbf{Unclear added value.}
Modeling is a design activity \cite{kendall2002systems}: it produces an artifact that is not yet a fully worked-out solution, but provides the design for one. 
Developing an actual, full solution from a model  involves writing a certain, potentially large amount of software code.
While advanced tools (especially in MDE \cite{ciccozzi2018teach}) allow automating some of the programming  via code generation,  there typically remains a large portion of code to be written by developers.
%, due to the abstraction gap between models and code.

Students often question the added value of models \cite{paige2014bad} as  an additional artifact that takes  effort to be created and maintained.
We can apply our interpretative lens to analyse this concern, which seems particularly justified in an analytical mindset:
There is no essential reason why models are \textit{required} for building a software system.
The system itself consists of code that provides the required functionality.
A complexivist mindset acknowledges the  effort invested into modeling, but, in turn, draws attention to the effort that is \textit{saved} from the use of models, in the context of realistic development settings:
Communication problems, misunderstandings and design flaws can be discovered early, before the system is built, which can help avoid significantly greater effort and time for fixing problems after the system has been built.
%In addition, newcomers have a clear entry point for understanding how the solution works.
To support development of this mindset, educational activities should be designed in a way that makes it likely for students to encounter situations that allow them to appreciate the advantages of modeling, e.g., in terms of saved effort.

\smallskip
\noindent{}\textbf{Understanding abstraction.}
Abstraction is an inherent principle of modeling:
To support analysis and design, models are created on an higher abstraction level than code, focusing on aspects of interest while intentionally removing details currently deemed outside of interest.
For example, a UML \textit{sequence diagram} focuses on the interaction of external actors and system components in terms of message exchanges, while abstracting away from all questions of how the components come up with the messages they send to the outside world.

Students regularly struggle with abstraction, and, consequently, deem modeling as too far away from programming \cite{brosch2009teaching,ciccozzi2018teach}.
These observations can be explained with the analytical-mindset-oriented tasks and problems that students consider in their pre-modeling education:
small-scale programming problems do not evoke a need for abstraction; it is sufficient to have a fully-worked out solution with all details in one place.
In fairness, such problems and the associated mindsets are  adequate for their educational purpose: for the “first steps” into computing, it is effective to deal with simple problems, which encourage a mindset focused on solving such problems. 
Yet, at some point, students need to deal with the complexity of real-world software. 
In a complevixist mindset, abstraction is a necessary means to that end:
The behavior of a system arises from the interaction of its components.
In the crucial notion of \textit{levels of emergence} \cite{davis2005challenging}, phenomena need to be understood at the level of their emergence, and cannot be understood in terms of lower-level, more detailed activities. 
For modeling education, this implies that, for students to understand the need for abstraction, we need to expose them to problems that cannot be solved without it.

\smallskip
\noindent{}\textbf{Putting things in context.}
Since modeling involves the use of modeling languages, learning to model also entails learning a new language.
As such, there is a tendency in courses to put an emphasis on understanding particular languages and associated tools.
For example, the typical first modeling language that students encounter is UML \cite{seidl2012uml}.
Much effort in such courses is invested into learning to use the language, typically by building small examples illustrating the main concepts.

With the focus on modeling languages, there is a tendency in courses to lack  guidance regarding the context in which modeling is performed, specifically: the problems that modeling tries to solve, the scope within the real world in which these problems occur, and the positioning of modeling in the   development process \cite{paige2014bad}.
In an analytical mindset, it is tempting to view this as a non-issue.
Following the reductionist approach, problems are given in the form of an exact  specification and problem solving entails splitting  the problem in smaller sub-problems, solving each sub-problem independently, by using  available problem-solving tools.
A complexivist mindset acknowledges the importance of learning in context.
Software addresses particular problems whose exact formulation my not be clear to begin with, and different involved actors, including the potential users, could have disagreeing views and interests in the solution.
Solving it might require deep understanding of the application domain, which might not necessary be one in which the developer has expertise in.
A complexivist mindset naturally encourages the use of project-based learning \cite{kokotsaki2016project}, which provides opportunities to understand a phenomenon in a rich, realistic context.

\smallskip
\noindent{}\textbf{Tool-related challenges.}
Modeling education typically encourages the use of industry-grade modeling tools \cite{paige2014bad}.
Arguably, using the same tools as professionals contributes to authenticity and allows students to benefit from useful  features.
In addition, students can benefit from useful tool functionality, either directly (e.g., support for layouting and automated feedback through validity checks) or  in the long run (e.g., collaboration and versioning support).

However, students  struggle with available modeling tools \cite{ciccozzi2018teach,batory2017teaching,paige2014bad}.
Their learning experience is affected by the perceived complexity and lack of user-friendliness in these tools, leading to decreased motivation.
From an analytical mindset, it is tempting to discourage the use of realistic tools in the first place:
as a problem-solving device, it would be equally valid to create models using  PowerPoint or  on paper;  factors that add accidental complexity \cite{brooks1987no} could be seen as noise and should be discouraged in education.
From a complexitivist mindset, using authentic tools seems generally  welcome, as it allows students to explore modeling in a realistic context, with all of the associated benefits and drawbacks.
However, the degree of realism needs to be managed, mirroring some concerns that the influence of allowing unrestricted complexity in education might not necessarily be positive \cite{hardman2011complexity}.
It is the teacher's responsibility to ensure that accidental tool issues do not get in the way with learning and forming a deeper understanding.
This theme bears a connection to the issue of facilitating independent problem solving when learning under guidance, which is explored in Vygotsky's idea of the \textit{Zone of Proximal Development}, and the notion of \textit{scaffolding} \cite{vygotsky1978mind}.

\section{Recommendations}
\label{sec:recommendations}

I now present  recommendations to address complexity in modeling education.
As put forward in Section~\ref{sec:tenets}, their underlying principle is to facilitate the adoption of a complexivist mindset, preparing the students for complex problems they will encounter in the real world, and explaining the role of modeling as a tool in this context.
Where I draw from cases from other disciplines, I reference them accordingly.
%This advice is not intended as \textit{one-size-fits-all} solution---I present it as ``building blocks'' that can be integrated in various course designs.
%In line with the principles of complexity thinking, I suggest to tailor the recommendations to the context at hand.
%For example, to offer a rich, realistic learning scenario, it is important to choose a scenario in which domain expertise is available, which depends on the teacher's background and network.

\smallskip
\noindent{}\textbf{Taking time to size up the system} is a recommendation lifted from  Gormley and Fenwick~\cite{gormley2016learning}, who observed that students performed better and were less prone to distractions when they took time to familiarize themselves with the environment before the exercise.
In modeling education, sources of complexity are tools, application domains, and the addressed problems.
Teachers can minimize the detrimental impact of accidental complexity inherent to tools by providing high-quality tutorials, which allows students to build a solid mental model for the tools during their first steps of using them.
In addition, for the considered application domain, they should provide students with domain expertise in a high-quality form, through descriptive text and media, as well as access to domain experts.
Teachers can support students during this process by dedicated tasks for exploring the tools and application domain, and by starting with simple, tutorial-like problems before moving to more ambitious ones (see the following item).

\smallskip
\noindent{}\textbf{Iterative and incremental teaching} is based on a practice by Fabricatore and López \cite{fabricatore2014complexity}, who apply a spiral process for studying  concepts and frameworks at increasing level of depth, allowing  understanding to increasingly deepen.
This advice transfers to modeling education in a direct way, given the complex nature of modeling languages of tools.
Educators should be mindful about the use of industry-grade languages and tools in beginner-level education.
When expected to work with them, students should be provided with clear usage explanations, possibly by pointing them to the particular parts of the tools to be used in assignments (usually a small subset).
Considering learnability- and usability-oriented languages \cite{acrectoaie2018vmtl,struber2017henshin}, or tools with a beginner mode (e.g., MagicDraw) is a viable alternative.
Tasks should be chosen in such a way that they naturally increase over time in difficulty and required expertise level.

\smallskip
\noindent{}\textbf{Offering a rich, realistic learning scenario}
to enhance authenticity  is a common theme in the considered studies, e.g., dealing with patient behavior \cite{gormley2016learning}, having a real stakeholder \cite{fabricatore2014complexity}, or emulating realistic projects \cite{thomas2008preparing}.
In modeling education, the associated advice is to move from  toy examples to realistic scenarios in meaningful  application domains.
To be able to give meaningful task descriptions and feedback, the lecturer should either be an expert (e.g., from developing relevant applications \cite{priefer2021applying}) or have access to experts in the application domain, possibly in the form of an actual stakeholder.
Of  interest are scenarios that span multiple disciplines,
%which would allow members of student groups to specialize on different disciplines,
mirroring Fabricatore and Lopez' advice to foster the emergence of specialisation \cite{fabricatore2014complexity}.
An example domain is robotics  \cite{garcia2020robotics}, which combines expertise from hardware, physics, navigation, and planning. 
%The scenario could be in a fast-moving field, e.g.. in 2023, it could include aspects of text generation with AI, which would then also be conducive to increasing the student interest
Choosing the right scenario requires deliberate navigation to reach an appropriate amount of complexity.
Replicating the full complexity of modern software systems in a typical university setting is, arguably infeasible, due to resource constraints and limited knowledge of educators and students \cite{sommerville2012large}.
Still, the cases from other disciplines  discussed before suggest that a “sweet spot” between toy problems and full-fledged real-world complexity is  attainable.

\smallskip
\noindent{}\textbf{Perturbations triggering adaptive dynamics} is based on a recommendation from Fabricatore and López \cite{fabricatore2014complexity}, who used consultant feedback and evaluations as a means to expose game design students to frequent and often unexpected pertubations, which required the groups to adapt.
In the case of modeling education, this strategy seems useful as well for exposing students to the intricateness of real-world software projects, where requirements can be vague and, especially in the case of agile development \cite{martin2003agile}, subject to frequent changes.
Pertubations could come from simulated stakeholders, who repeatedly change their needs, or changes between successive assignments. Descriptions can contain deliberately vagueness and contradictions that become apparent during modeling.
Planned pertubations require a certain degree of moderation, as students may feel cheated when they are faced with changing assignments in an educational context.
At the same time, there is a risk of overexplaining.
%as providing context and explanations for every instance of planned perturbations would affect authenticity.
 I recommend to have several general disclaimers in course materials and early course meetings that the coursework might involve changing and unclear requirements, explaining the educational purpose.

\smallskip
\noindent{}\textbf{Planned non-linearity}, a new recommendation informed by the very definition of complex systems, acknowledges that complex software is non-linear.
Non-linearity arises when new features are developed that can potentially interact with all other existing features, leading to the so-called \textit{feature interactions} \cite{cameron1993feature}. 
%For example, a phone operating system might have features for auto-forwarding a call, and for automatically putting callers on hold---if both features are present, the desired behavior needs to be specified somewhere.
Another source of non-linearity are \textit{feedback loops} in software with machine learning components \cite{sculley2015hidden}, in which decisions made by a machine learning model can influence the system's outcomes and leak into newly included training data, eventually leading to unintentional amplification of signals from the initial training data.
Including  sources of non-linearity in  problem descriptions leads to complex programming problems, where models can help students to gain awareness and a better understanding of interactions and feedback loops, for example, by visualizing dependencies in \textit{component diagrams}.

% \label{sec:conclusion}
% Studying the educational practice from other disciplines can be insightful to inform the teaching of modeling.
% The most important direction for future work is to  study the impact of complexity in modeling education empirically.

\section{Conclusion}
\label{sec:conclusion}
Studying the educational practice from other disciplines can be insightful to inform the teaching of modeling.
The most important direction for future work is to  study the impact of complexity in modeling education empirically.

I revisited recurring challenges in modeling education from the lens of complexity science, and offered recommendations for improved teaching practices.
Drawing from complexity-science-inspired strategies in the educational practice of four other disciplines, I presented five recommendations with the common goal of supporting the adoption of a complexivist mindsets by students.

A future work, first, I envision
%as a follow-up to this conceptual analysis, it seems worthwhile to conduct an 
an a empirical study of the usefulness of the proposed recommendations, studying short-term implications   on course outcomes, and long-term effects based on feedback from students who moved on to developer roles.
Second, a more comprehensive study of complexity-science-based teaching practices in various disciplines and their potential impact to modeling education.
%While considering the four disciplines from the literature review was already insightful,  a saturation point might not have been reached.

\section*{Acknowledgement}
The author wishes to thank the teachers and participants from Gothenburg University's  course \textit{Teaching and Learning in Higher Education 3: Applied Analysis (PIL103)}, especially  Janna Meyer-Beining, for high-quality feedback and insightful discussions on previous versions of this article.

\bibliographystyle{ieeetr}
\bibliography{main}

\end{document}